\documentclass[journal=jacsat,manuscript=article]{achemso}
\usepackage[version=3]{mhchem} % Formula subscripts using \ce{}
\usepackage{booktabs}
\usepackage{color}
\usepackage{amssymb}

\newcommand{\degree}{$^{\circ}\hspace{4px}$}
\newcommand{\ang}{\text{\normalfont\AA}}
\newcommand{\arcangle}{%
  \mathord{<\mspace{-9mu}\mathrel{)}\mspace{2mu}}%
}
\author{Vladim\'ir Zoba\v{c}}
\email{vladimir_zobac001@ehu.eus}
\affiliation[CFM]
{Centro de F{\'i}sica de Materiales CFM/MPC (CSIC-UPV/EHU)\\
Paseo Manuel de Lardizabal 5, 
E-20018 Donostia - San Sebasti\'an, Spain.}
\author{Roberto Robles}
\affiliation[CFM]
{Centro de F{\'i}sica de Materiales CFM/MPC (CSIC-UPV/EHU)\\
Paseo Manuel de Lardizabal 5, 
E-20018 Donostia - San Sebasti\'an, Spain.}
\author{Nicol\'as Lorente}
\affiliation[CFM]
{Centro de F{\'i}sica de Materiales CFM/MPC (CSIC-UPV/EHU)\\
Paseo Manuel de Lardizabal 5, 
E-20018 Donostia - San Sebasti\'an, Spain.}
\alsoaffiliation[DIPC]
{Donostia International Physics Center (DIPC)\\
Paseo Manuel de Lardizabal 4, 
E-20018 Donostia - San Sebasti\'an, Spain.}

\title[ABP] {Directionality in van der Waals Interactions: the Case of 4-Acetylbiphenyl Adsorbed on Au(111)}
\abbreviations{DFT}

\begin{document}

\begin{tocentry}

\begin{center}
\includegraphics[scale=1.0]{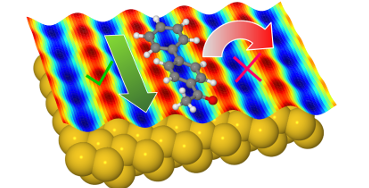}
\end{center}

\end{tocentry}

\begin{abstract}
We report on a theoretical study of adsorption of 4-Acetylbiphenyl molecule and its 
diffusion properties in the main directions of the Au(111) surface.  Structural changes 
of the molecule, which are induced by adsorption lead to stronger conjugation of the 
$\pi$-system. The molecule is adsorbed in a flat configuration on the surface
with roughly the same binding energy along the [1\={1}0] and [11\=2] directions, in
good agreement with experiments. Furthermore, the diffusion barriers
imply an important directionality of the molecule-surface interactions. This
is somewhat surprising because our calculations show that the prevailing
interaction is the long-range molecule-surface van der Waals interaction.
Despite of its weakness, the van der Waals interaction discriminates
the preferential adsorption sites as well as imposes a molecular
geometry that needs to be considered when rationalizing the diffusion barriers.
\end{abstract}

\section{Introduction}

New materials based on metal-organic interfaces offer promising properties
 in many applications concerning photovoltaic devices, sensors
or heterogeneous catalysis \cite{Clair2019, Tsutsui2012, Giovanni2010,
Rosei2003}.  The overlayer properties can be tuned by the nature of
adsorbates and surface, as well as  the way they  assemble.

An important strategy to understand the interactions at play in the
creation of molecular overlayers is atomic manipulation with scanning
probes. The way molecules displace, assemble or dissociate under the
action of such probes gives direct information of the forces joining them
together and the role of the environment in the formation of collective
structures.  The various manipulation strategies include pushing molecules
by chemical or electrostatic interactions with the STM tip, and excitation
by tunneling electrons~\cite{Morgenstern2013,Eisenhut2018, Nickel2013,
Ohmann2015, Simpson2017}.

A basic ingredient to understand the movement of molecules
is the evaluation of the molecular adsorption positions,
geometries, and potential energy surfaces~\cite{Zhang2016, Liu2012,
Pekoz2016,Martinez2012,Meyer2004}.  A particularly important aspect of
molecular motion is the understanding of diffusion barriers along the
holding surface.

In this work, we study the adsorption of the 4-acetylbiphenyl
(ABP) molecule on Au(111).  The present calculations are motivated
by recent experiments where ABP is shown to be an excellent
building block for supramolecular structures that show such a
strong individual character that can be manipulated as single
objects~\cite{Nickel2013,Ohmann2015,Eisenhut2016} at liquid He
temperature. In parallel, the weakness of the intermolecular bonding
allows ABP to form different structures~\cite{Eisenhut2016}.  The main
data~\cite{Nickel2013,Ohmann2015,Eisenhut2016} show that more than 90\%
of the molecules can be moved along the surface [1\={1}0] direction.
The rest of the motion involves other directions or a combined motion of
rotation and translation. The displacements along the [1\={1}0] direction
are 60\% in steps of a single lattice unit, and 40\% of two lattice
units. Very rarely larger steps are found.  A defining experimental
fact of the different structures is that the molecules adsorb flat on
the surface either along the [1\={1}0] direction or the [11\=2] one. No
other possibilities are found.  This shows an important directionality
and selectivity of the ABP-surface interaction that is important to
unravel for the understanding of supramolecules and overlayers based on
biphenyl derivates.

\section{Methods}

All calculations were performed with the Vienna ab initio simulation package
(VASP) using the PBE approximation to density functional theory
\cite{KRESSE199615,PBE}. 
Valence electrons were described using a plane-wave basis
 with an energy cutoff of 400 eV  and 
projected-augmented-wave potentials \cite{Kresse1999}. 

We used the Tkatchenko-Scheffler method \cite{Tkatchenko2009} for the
description of the long-range van der Waals (vdW)  interactions as
implemented in VASP.  This method has been shown to reliably describe
the adsorption properties of organic molecules on noble-metal surfaces
\cite{Tkatchenko_2016,Laura}. It consists in the use of free-atomic
polarizability to compute the coefficient multiplying the $1/R^6$
behavior, together with a damping function that allows the recovery of the
PBE interactions at short range. Here, the Tkatchenko-Scheffler method
is applied non-self-consistently such that the dynamical polarization
effects are the ones of the PBE functional.

To simulate the surface, we used a Au (111) $9\times5\sqrt{3}$
slab with 4 atomic layers with an inter-slab separation of 20
$\textup{\AA}$.  Structure relaxations were performed until the
atomic forces on the molecule and first two layers were smaller than
$0.02~\textup{eV/\AA}$. The k-point samplings were $3 \times 3 \times 1$
regular grid of k-points \cite{Monk}. The coarse PES (here, used for
illustrative purposes) were obtained  using a smaller unit cell with a
$8\times4\sqrt{3}$ slab with 3 atomic layers and one k-point. In order
to calculate PES the molecule was systematically placed over the one
unit cell with grid sampling $10\times10$ points. In this case the surface
and molecule were fixed and single energy point calculations were
performed. The total energy was converged below $10^{-6}$ eV in both
cases. The obtained  Au bulk lattice constant under the above conditions
is 4.114 \AA .

We have also used the
nudged elastic band method (NEB method)\cite{NEB} 
to compute the minimum-energy path between two conformations on
the surface.

The method works by setting an elastic constraint between intermediate
conformations and simultaneously allowing them to relax until the minimum
energy is attained. The molecule in each position is relaxed but keeps
appropriate distance between neighbouring intermediate states (or images). In this way, the
obtained energies give the minimum energy path. A critical aspect of
the NEB method is the determination of the intermediate states. 
These images were first calculated  by
a linear interpolation between initial and final states, and
then allowed to relaxed following the elastic-band constraint.
Including the initial and final
states, 6 images were sufficient to sample the diffusion pathway.

\section{Results and Discussion}

%\section{Adsorption of  4-acetylbiphenyl on Au (111) }
\subsection{The 4-acetylbiphenyl molecule in gas phase}

The 4-acetylbiphenyl molecule (ABP) is formed by two isolated phenyl
rings and acetyl group. The phenyl rings are joined by a single
C-C bond that permits the rings to mutually rotate.  The molecule
presents an angle between phenyl rings due to the tendency to flatten
the molecule in order to optimize the $\pi$-conjugation between
the rings, and the counterbalancing of steric hindrance between
adjacent hydrogens~\cite{biphenyl}.  Our calculations yield an angle
of 35.8\degree. This is 3.85\degree smaller than the one found for the very similar compound
biphenyl-4-carboxaldehyde using quantum chemical methods to reproduce the experimental
data~\cite{SRISHAILAM}. The difference with the present calculation may
be caused by the presence of an aldehyde group instead of the acetyl group in our
case. 

\begin{figure}
  \includegraphics[scale=0.7]{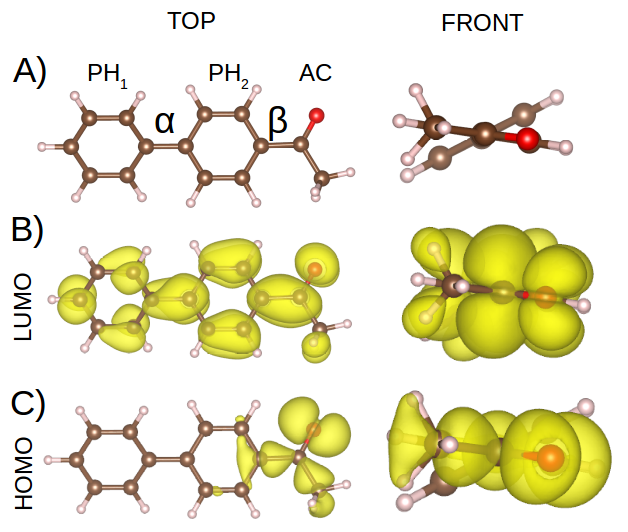}
  \caption{Geometrical structure of ABP molecule A), LUMO B) and HOMO C) orbitals of the ABP molecule.}
  \label{gas}
\end{figure}

Our calculations confirm the tendency of the phenyl rings to
significantly separate from a planar configuration.  Figure~\ref{gas}
(A) shows the relaxed molecular structure. The acetyl group replaces the
hydrogen atom at the fourth carbon atom in one of the phenyl rigns. As a
consequence, the molecule develops a small dipole of 1.455 Debyes. The
presence of the acetyl group makes possible the formation of hydrogen
bonds with similar molecules.

Figure~\ref{gas} (B) and (C) shows the frontier orbitals of ABP. The
highest occupied molecular orbital (HOMO) is a sigma orbital largely
concentrated on the acetyl group.  It contains an important contribution
from the 2$p_x$ orbital of the oxygen atom (taking the $z$ coordinate
as perpendicular to the surface and the $x$ one along the p-orbital).
This feature makes it particularly important to describe the interactions
where the O atom is involved. The lowest unoccupied molecular orbital
(LUMO) is in contrast a delocalized $\pi$ orbital extended over the
full molecule.

\subsection{The adsorbed molecule}

\begin{figure}
  \includegraphics[scale=0.6]{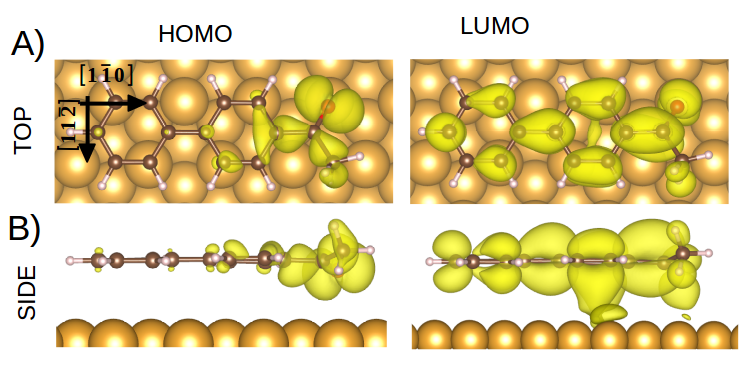}
  \caption{Geometrical structure of ABP molecule adsorbed along the surface
[1\=10] direction. The yellow contour depicts the shapes of
two orbitals of the system that largely match the HOMO and LUMO
of the free molecule. 
The small distortion of the LUMO shows the smallness of covalent-like interactions between molecule and surface.
Isosurface of the plot is 0.01.}
  \label{fig:str_110}
\end{figure}

\begin{figure}
  \includegraphics[scale=0.6]{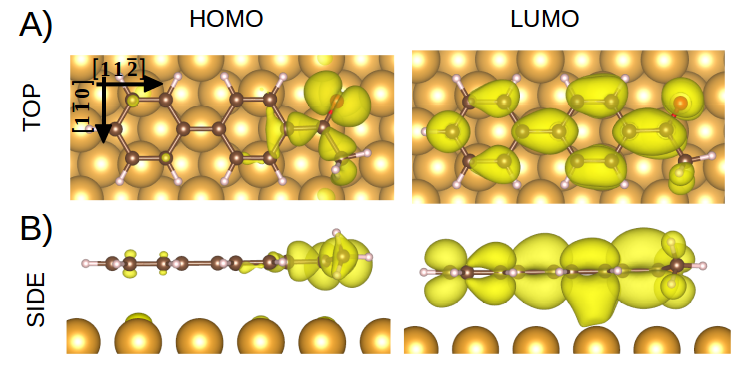}
  \caption{Same as figure~\ref{fig:str_110} for a
 ABP molecule adsorbed along the surface
[11\=2] direction.}
  \label{fig:str_112}
\end{figure}

Following the experimental findings of Ref.~\cite{Nickel2013}, we position
the molecule in two different conformations. The first one consists in
aligning the two phenyl rings with the surface [1\={1}0] direction and
the second one with the [11\={2}] direction. The optimization of the
two structures entails full relaxation and exploration of the lowest binding
energy along the surface unit cell. As is usually the case for aromatic
molecules on Au (111) (see for example Ref.~\cite{Tkatchenko_2016}),
the phenyl rings optimize their overlap with the surface hollow sites.
Figures~\ref{fig:str_110} and \ref{fig:str_112} show the minimum energy
conformations.

\begin{table}
 \caption{Table of binding energies and geometrical properties of ABP molecule. Distance of C$^n_{aver}$-Au[\ang] was calculated as a difference of z coordinate averaged over all carbon atoms and z coordinate of first atomic layer.  }
  \centering
  \begin{tabular}{ |c|c|c|c|c|c|}
    \hline
%    \multicolumn{2}{c}{Part}                   \\
%    \cmidrule(r){1-2}
    Structure     & E$_{ads}$ [eV] & dist. O-Au [\ang] & dist. C$^n_{aver}$-Au[\ang] & $\arcangle$ $\alpha$ [$^{\circ}$]  & $\arcangle$ $\beta$ [$^{\circ}$]   \\
    \hline
    ABP$^{free}$        &    ---       & ---    & ---  & 35.80 & 1.40    \\
%    \hline
    ABP$^{1\bar{1}0}$    &   -1.602       & 2.86    & 3.28  & 3.10  & 15.20    \\
%    \hline
    ABP$^{11\bar{2}}$    &    -1.588       & 3.16    & 3.37  & 0.80   & 6.70   \\
    \hline
  \end{tabular}
  \label{tab:t_B_en_geom1}
\end{table}

The largest molecular change induced by adsorption is the flattening of
the inter-ring angle. The van der Waals interaction is optimized for a large overlap
of the ring with the surface, reducing the inter-ring $\alpha$ angle 
from 36\degree to 3\degree for the [1\={1}0] configuration and
1\degree for the [11\=2], see Table~\ref{tab:t_B_en_geom1}.
This leads to a stronger conjugation of the
$\pi$-system induced by the adsorption on the surface.

The binding energy, Table~\ref{tab:t_B_en_geom1}, is defined by
$E_{ads}=E_{surf+mol}-(E_{surf}+E_{mol})$. The [1\=10] conformation is
14 meV lower than the [11\=2] one. The binding energy is rather
strong around $-1.6$~eV, very close to the values found for
naphtalene and related organic molecules on densely packed noble
metal surfaces~\cite{Tkatchenko_2016,Laura}. Moreover, the average height
of the molecule to the substrate (C$^n_{aver}$-Au, Table~\ref{tab:t_B_en_geom1})
is in the same range of distances~\cite{Tkatchenko_2016,Laura}.
As in all those cases, the interaction of ABP with Au (111) is
given by the vdW interaction between molecule and substrate.
Other types of interactions are largely absent.

In addition to the small energy difference between the [1\=10] and the [11\=2]
conformations, the acetyl group present different angles (see entry $\beta$ in
Table~\ref{tab:t_B_en_geom1}). For the [1\=10] conformation the O atom is significantly 
closer to the surface leading to an acetyl angle twice as large. These two
differences are connected and show the interaction of the O atom with
one of the top Au atoms of the surface. Two
states reminiscent of the LUMO and HOMO are plotted in
figures~\ref{fig:str_110} and \ref{fig:str_112}. They show very small differences compared to the molecule in the gas phase (Fig.~\ref{gas}) due to the very weak O-Au interaction.

\begin{figure}
  \includegraphics[scale=0.5]{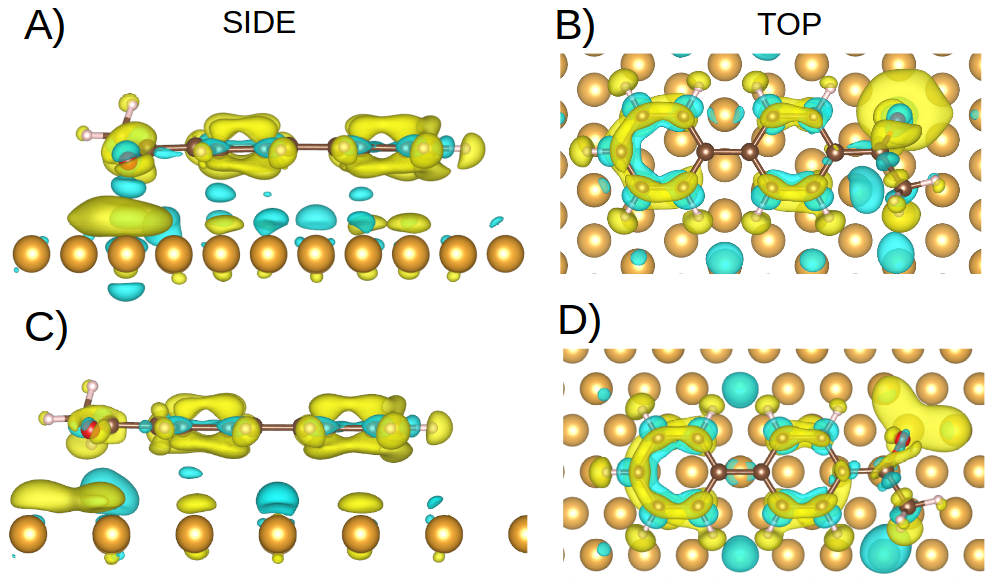} 
  \caption{Induced density of ABP in [1\=10] A),B) and ABP in [11\=2] C),D). The main
difference involves the induced density about the oxygen atom and the
area underneath the oxygen atom. The isosurface of the plot is 0.0005 $e/$\AA$^3$~and yellow and green colors denote positive and negative values respectively.}
  \label{fig:I_d_110_112}
\end{figure}

Charge transfer between molecule and surface is also absent. Bader
analysis~\cite{Bader} leads to less than 0.1 electrons
of reduction of the molecular electrons. As a consequence, all interatomic
distances within the molecule are identical to the gas phase ones within
numerical accuracy. 
Despite the absence of charge transfer, there is a polarization of the
electronic clouds locally around molecule and substrate, that
point at the importance of
vdW interactions in this system, figure~\ref{fig:I_d_110_112}. The
induced density is defined as the difference of electronic density between the adsorbed system and the sum
of the surface and molecular densities keeping the same geometry
but separately computed. As expected, the differences between both
conformations reduces to the difference in oxygen
position with respect to the surface.

\begin{figure}
  \includegraphics[scale=0.5]{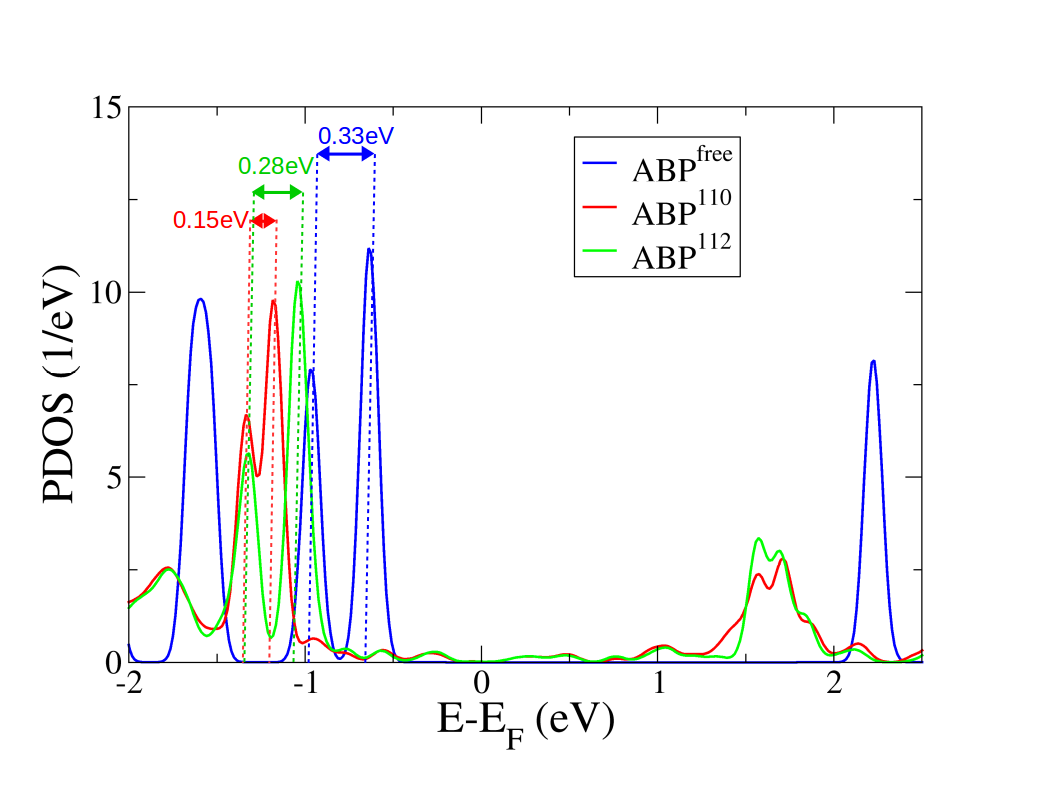} 
  \caption{Density of states projected on the orbitals of the ABP molecule in 
the gas phase (marked as ``free'' and plotted in blue color) and
the adsorbed molecule along
the [1\=10] direction (red color) and along the [11\=2] one (green color).
The vacuum level of the surface and gas-phase systems have been aligned. This shows
a sizable downshift in energies with modification of the density of states
due to the interaction between molecule and substrate.}
  \label{fig:DOS_100_112_vac}
\end{figure}

The small overlap of the LUMO with surface states in figures~\ref{fig:str_110}
and \ref{fig:str_112} does explain the 0.5 eV downward shifting and
broadening of
the LUMO in the projected density of states (PDOS), see
figure~\ref{fig:DOS_100_112_vac}. 
The blue line corresponds to the PDOS of the gas-phase molecule, while the
red and green refer to the PDOS of the [1\=10] and [11\=2] conformations.
As gleaned from figures~\ref{fig:str_110}
and \ref{fig:str_112} the overlaps between the LUMOs of both configurations and the substrate are the same leading to equal PDOS. 
However, the behavior of the HOMO is more intricate. 

The molecular orbital structures on the surface are determined by the
molecule-surface interactions and referred to the surface Fermi energy dividing between occupied and empty electronic states. The gas-phase molecular structure is aligned to the ones on the surface using their respective vacuum levels.
The HOMO contribution is the first occupied peak, just below the Fermi level
(the zero
of energies in figure~\ref{fig:DOS_100_112_vac}). We see that the gas-phase
HOMO coincides with the HOMO for the [11\=2] conformation, and that the HOMO --
HOMO-1 distance is maintained. However, the [1\=10] conformation
shows a downward shift of 0.5 eV and a closing of the HOMO -- HOMO-1 distance.
We rationalize these shifts by the larger interaction of the O atom with
the surface for the [1\=10] case as described above. Indeed, the important
contribution of the O $p$-orbital to the HOMO, makes it shift due to the
interactions caused by the O atom. These interactions are smaller or even absent 
in the other orbitals where the relative weight of the O atom is lesser.
Nevertheless, the bonding analysis is better performed in terms of
the electronic density and its derivatives instead of the molecular
orbitals. A clear example of this is given by the NCIplot methodology
applied to molecules~\cite{Johnson2010}, periodic systems~\cite{Otero2012} and the
organic-metal surface interface~\cite{Boto2015}.

\subsection{Diffusion barriers along the Au (111) surface}

Ample experimental results~\cite{Nickel2013} show that ABP diffuses
on the Au (111) surface.  The experiments report on the formation of
supramolecular structures~\cite{Nickel2013} that themselves can be
controllably moved by applying electronic pulses from an STM. Study
of these manipulations show that the ABP molecules preferentially move
along the high-symmetry directions of the surface and in small integer
steps of the surface lattice parameter.  These findings are unexpected
for a vdW-bonded molecule, because the vdW interaction shows a very small
corrugation with surface sites. Rather, a uniform potential energy surface
(PES) is expected with barriers that are negligible compared to the
overall binding energy.  It is interesting then to compute and rationalize
the barriers that ABP encounters when diffusing along the surface.

\begin{figure}
  \includegraphics[scale=0.4]{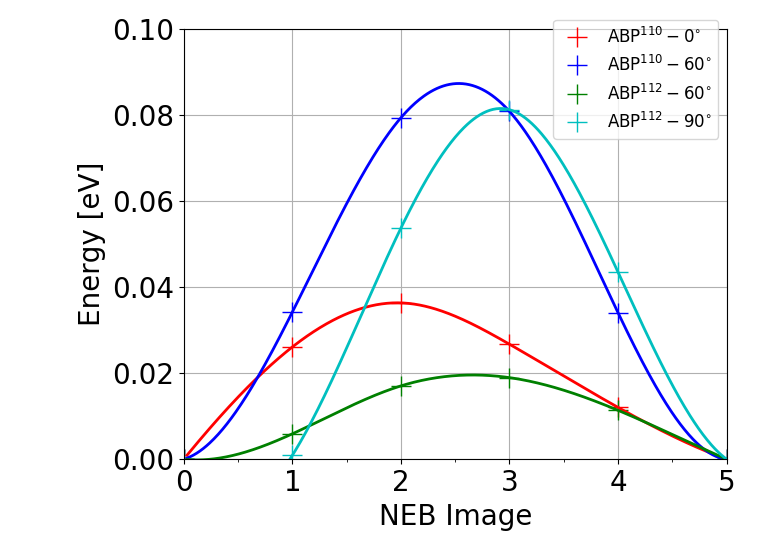} 
  \caption{Chosen directions along the  adibatic potential energy
surface calculated by NEB for a molecule adsorbed along the [1\=10] direction
(ABP$^{1\bar{1}0}$) and along the [11\=2] direction (ABP$^{11\bar{2}}$). For the ABP$^{1\bar{1}0}$
case the molecule is moved in one of the 3 equivalent [1\=10] directions of the Au
(111) surface. For the  0\degree case the molecule follows the same direction
it points at (red line). For the 60\degree one, the molecule follows a [1\=10] direction
that makes a 60\degree angle with the direction of the molecule (blue line). The
ABP$^{11\bar{2}}$ molecule along a 60\degree path finds a very small barrier (green
line). However, the
ABP$^{11\bar{2}}$ molecule founds the maximum barrier when it is moved along
a [1\=10] direction, the 90\degree case (cyan line). Energy points of the barriers were fitted by cubic spline.
  \label{fig:NEB}}
\end{figure}

Figure~\ref{fig:NEB} shows the adiabatic PES along four directions. The
first two (red and blue) are for a molecule adsorbed along the [1\=10]
direction. The red curve is a path following the [1\=10] direction when
0 \degree is between molecule axis and path as marked on the figure
\ref{fig:str_110}. We see that 5 images suffice to yield a smooth
representation of the PES. The barrier is less than 40 meV. Compared to
the 1.6 eV of binding energy of the ABP molecule with [1\=10] orientation,
we can conclude that the barrier is indeed small. The blue path presents
a stronger barrier of roughly 80 meV. Again, this is a small corrugation
of the PES.

The blue path is for a molecule moving along a [1\=10] direction rotated
60 \degree with respect to the original [1\=10] direction. The barrier is
much higher because the molecule is not aligned with the moving direction.
Table~\ref{tab:EN_barriers} gives the barrier values and the contribution
of vdW interactions to the barrier (E$^{disp}$$_{barrier}$). We can
see that the vdW interaction alone accounts for the barrier. The 80-meV
barrier is then the corrugation of vdW landscape of the ABP molecule on Au
(111).

When the molecule is adsorbed along the [11\=2] direction,
figure~\ref{fig:str_112}, the barriers also depend on what [11\=2] direction is
taken with respect to the molecule, figure~\ref{fig:NEB}, cyan and green
curves. The vdW corrugation suffices to explain the barriers. Only when
the barriers are small, do other corrugations play a role. Indeed, the
small changes in conformation of the molecule as it is pushed above the
barrier explain the contributions to the barrier that are not 100\% vdW.

\begin{table}
 \caption{Table of energy barriers.}
  \centering
  \begin{tabular}{ |c|c|c|}
    \hline
    Structure/direction     & E$^{tot}_{barrier}$ [eV] & E$^{disp}_{barrier}$ [eV]   \\
    \hline
    ABP$^{1\bar{1}0}$-0$^{\circ}$    &    0.036       & 0.023  \\
%    \hline
    ABP$^{1\bar{1}0}$-60$^{\circ}$    &   0.087       & 0.080   \\
%    \hline
    ABP$^{11\bar{2}}$-60$^{\circ}$    &    0.019       & 0.009   \\
%    \hline
    ABP$^{11\bar{2}}$-90$^{\circ}$    &    0.081       & 0.070   \\
    \hline
  \end{tabular}
  \label{tab:EN_barriers}
\end{table}

\begin{figure}
  \includegraphics[scale=0.6]{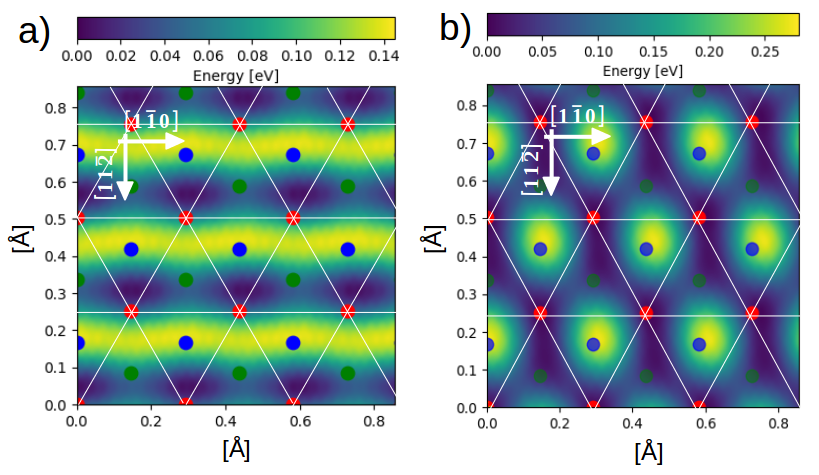} %{structure110.png}
  \caption{Potential energy surfaces of ABP with
  its axis along the [1\=10] direction a) and [11\=2] direction b) Red,
  blue, green dots indicate  position of Au atoms in first, second and
  third layer. White straight lines are an eye-guide for Au rows in
  first layer. In agreement with the NEB paths of figure~\ref{fig:NEB}, we
see that the barrier is minimum along the [1\=10] direction that aligns
with the molecular axis.}
  \label{fig:PES110_112}
  \end{figure}

A clearer representation of the above paths, figure~\ref{fig:NEB}, can
be seen in figure~\ref{fig:PES110_112} where the center of mass of the
molecule has been displaced along the surface and the internal coordinates
have been maintained in the minimum energy configuration. This approximate
representation is  intended for visualization purposes.
Now, the directions given
in figure~\ref{fig:NEB} can be easily pictured. The PES profiles allow
us to conclude that there is one preferred direction of diffusion for
each configuration, as well as a less preferred direction. 
These observations may play a role in the explanation
of the collective uniform diffusion of supramolecular structures composed
of ABP molecules.

\begin{figure}
  \includegraphics[scale=0.6]{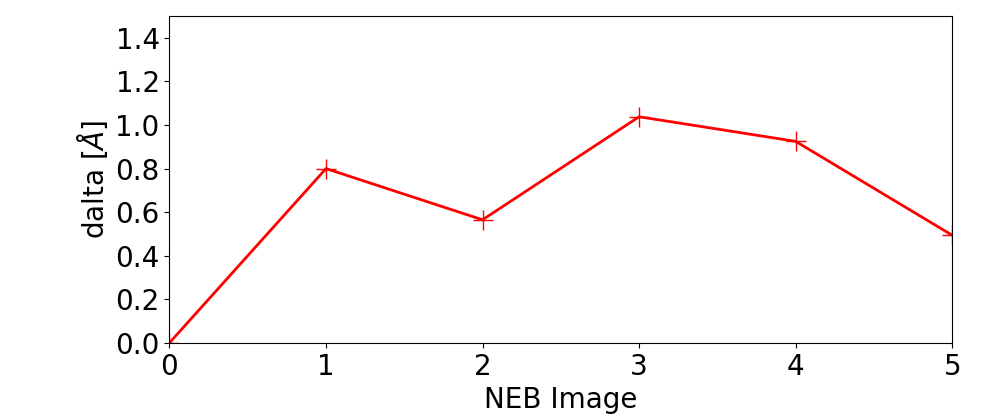}
 \caption{Accumulated displacement of the molecular atomic positions once
the center of mass has been shifted together. Molecule is moved in the [11\={2}] direction. The geometrical distortion is maximum at the transition state, in agreement with a geometrical
interpretation of the diffusion barrier.}
\label{fig:distortion}
  \end{figure}

A contribution to the
barrier and, hence, to the directionality of the interactions is
given by the distortion of the molecule.
In order to estimate the distortion, we have compared the nuclear positions
of the molecular atoms along the NEB path. In figure~\ref{fig:distortion} we compute
\begin{equation}
delta = \sum_{atom} |\vec{R}_{NEB}-\vec{R}_{relaxed}|
\end{equation}
where $\vec{R}_{NEB}$ are the coordinates of each atom for a given NEB image and $\vec{R}_{relaxed}$
are the coordinates for the relaxed molecules, where the center of mass has been shifted
to agree in the compared configurations. At the moment of the transition, we find
an accumulated displacement of 1~\AA~in the coordinates of the molecule, which amounts to
an avarage displacement of more than 0.05~\AA~per carbon atom. This goes in the
direction of the increase of energy as we move through the transition state.

\begin{figure}
  \includegraphics[scale=0.5]{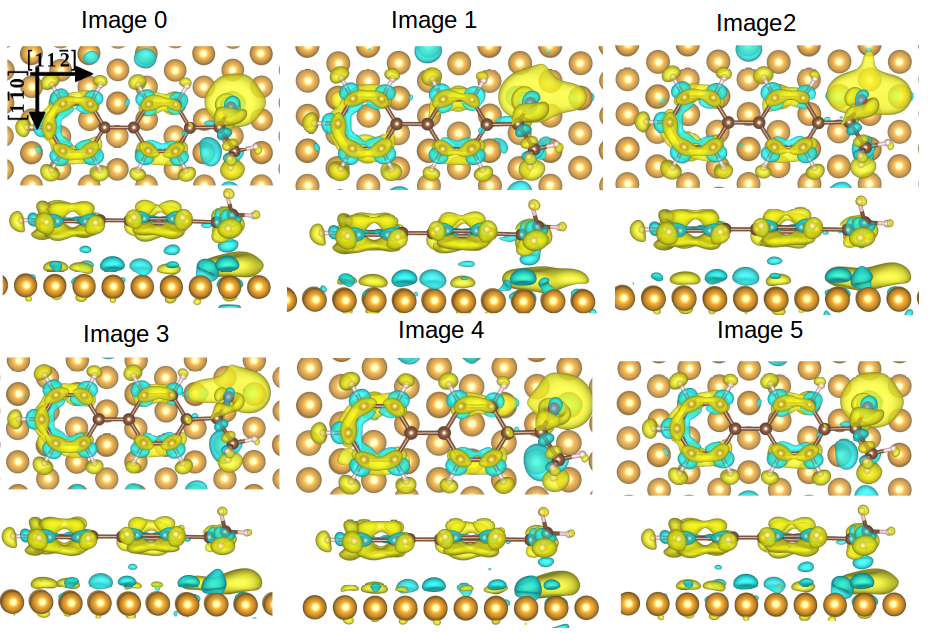}
 \caption{
Induced electronic density as the molecule conformation changes along
the different images of the minimum-energy path given by the NEB calculation. The molecule is translated in [11\=2] direction.
Largely the oxygen atom is polarized as it changes its registry with the surface. The
polarization of this calculation is the one given by the PBE functional.
}
\label{NEB_id}
  \end{figure}

An important contribution for the vdW interaction comes from the polarization of
the different components of the system. The induced electronic density permits
us to asses the change of the polarization as the molecule moves through
its diffusion path given by the NEB calculation. Figure~\ref{NEB_id} show the
change of induced density for the different images of the NEB calculation. We
see that the phenyl rings are little affected by polarization while the oxygen
atom suffers important perturbation as it moves over the surface,
particularly images 1 and 2. The polarization
is very local and it is mainly due to the Au--O interaction. The oxygen
atom is largely inert at the molecule but it is easily polarizable as the polarization
included in the PBE functional shows.  

These results show that simpler molecular dynamics methods with simple
potentials will fail to capture the large polarizability of the metallic
surface, where the PBE functional is adequated.

\section{Conclusions}

This study reveals the adsorption properties of 4-acetylbiphenyl on
Au (111) that are important to rationalize the atomic manipulations
performed with an STM in Ref.~\cite{Nickel2013}. As with other aromatic
molecules adsorbed on noble metal surfaces, van der Waals interactions
are the prevailing interactions. As a consequence, very small molecular
changes happen during adsorption, except for the flattening of the two
phenyl rings. Due to the small perturbation exerted by the surface on
the molecule, charge transfer is less than 0.1 electrons and molecular
orbitals are easily recognized on the adsorbed molecule.

The main experimental data~\cite{Nickel2013,Ohmann2015,Eisenhut2016} 
are that the molecules adsorb flat on the surface either along
the [1\={1}0] direction or the [11\=2] one and that
they diffuse (more than 90\%) along the [1\={1}0] direction in steps of one
or two surface-lattice parameter. 
These data
can be rationalized by the corrugation of the vdW interaction. Indeed, the vdW
interaction follow the corrugation of the surface, as a consequence
the molecule will follow steps matching units of the surface-lattice parameter.
The fact that only one or two steps are experimentally available means
that a sizeable damping of the motion is available probably
given to the continuum of electronic excitations available from zero energy
on a metallic surface.

Unfortunately, no experimental barrier has been obtained. The computed PES show
adiabatic barriers are below 90 meV. Experimentally, the molecule
is moved when bias pulses larger than 2 volts are applied. This means that electrons
from the STM tip need to reach 2 eV before some molecular dynamics can be
detected. However, this is not connected with the molecular energy barriers
but rather with the energy-dependence of the electron-molecule cross section. Our
density of state shows that the molecular states lie at biases in the eV range
facilitating the electron-molecular collision at eV energies. 

The computed barriers are negligible
in front of the binding energy of the surface (-1.6 eV). 
This corroborates the common
idea that van der Waals interactions lead to rather flat PES.
Our calculations further show that the intrinsic
dipole of the molecule is too small to play a role in the
adsorption or diffusion of the molecule.

This work has permitted us to rationalize in terms of the van der Waals
interaction that at a few K of temperature, the molecules 
stably adsorb, can assemble and be easily manipulated with a scanning probe,
as experimentally found~\cite{Nickel2013}.

\begin{acknowledgement}
We acknowledge the EU commission H2020 FETopen program \textit{Mechanics
with Molecules}, project number 766864.
We are also grateful for financial support from the Spanish
 MICINN, project RTI2018-097895-B-C44. Computer resources 
were obtained at the RES computers Finisterrae II
in project RES-QCM-2019-1-0024 and Cibeles in project RES-QS-2019-3-0012
and are gratefully acknowledged.

\end{acknowledgement}

\bibliography{zobac}

\providecommand{\latin}[1]{#1}
\makeatletter
\providecommand{\doi}
  {\begingroup\let\do\@makeother\dospecials
  \catcode`\{=1 \catcode`\}=2 \doi@aux}
\providecommand{\doi@aux}[1]{\endgroup\texttt{#1}}
\makeatother
\providecommand*\mcitethebibliography{\thebibliography}
\csname @ifundefined\endcsname{endmcitethebibliography}
  {\let\endmcitethebibliography\endthebibliography}{}
\begin{mcitethebibliography}{30}
\providecommand*\natexlab[1]{#1}
\providecommand*\mciteSetBstSublistMode[1]{}
\providecommand*\mciteSetBstMaxWidthForm[2]{}
\providecommand*\mciteBstWouldAddEndPuncttrue
  {\def\EndOfBibitem{\unskip.}}
\providecommand*\mciteBstWouldAddEndPunctfalse
  {\let\EndOfBibitem\relax}
\providecommand*\mciteSetBstMidEndSepPunct[3]{}
\providecommand*\mciteSetBstSublistLabelBeginEnd[3]{}
\providecommand*\EndOfBibitem{}
\mciteSetBstSublistMode{f}
\mciteSetBstMaxWidthForm{subitem}{(\alph{mcitesubitemcount})}
\mciteSetBstSublistLabelBeginEnd
  {\mcitemaxwidthsubitemform\space}
  {\relax}
  {\relax}

\bibitem[Clair and {De Oteyza}(2019)Clair, and {De Oteyza}]{Clair2019}
Clair,~S.; {De Oteyza},~D.~G. {Controlling a Chemical Coupling Reaction on a
  Surface: Tools and Strategies for On-Surface Synthesis}. \emph{Chemical
  Reviews} \textbf{2019}, \emph{119}, 4717--4776\relax
\mciteBstWouldAddEndPuncttrue
\mciteSetBstMidEndSepPunct{\mcitedefaultmidpunct}
{\mcitedefaultendpunct}{\mcitedefaultseppunct}\relax
\EndOfBibitem
\bibitem[Tsutsui and Taniguchi(2012)Tsutsui, and Taniguchi]{Tsutsui2012}
Tsutsui,~M.; Taniguchi,~M. {Single molecule electronics and devices.}
  \emph{Sensors (Basel, Switzerland)} \textbf{2012}, \emph{12}, 7259--98\relax
\mciteBstWouldAddEndPuncttrue
\mciteSetBstMidEndSepPunct{\mcitedefaultmidpunct}
{\mcitedefaultendpunct}{\mcitedefaultseppunct}\relax
\EndOfBibitem
\bibitem[Bottari \latin{et~al.}(2010)Bottari, de~la Torre, Guldi, and
  Torres]{Giovanni2010}
Bottari,~G.; de~la Torre,~G.; Guldi,~D.~M.; Torres,~T. {Covalent and
  Noncovalent Phthalocyanine-Carbon Nanostructure Systems Synthesis,
  Photoinduced Electron Transfer}. \emph{Chem Rev} \textbf{2010},
  6768--6816\relax
\mciteBstWouldAddEndPuncttrue
\mciteSetBstMidEndSepPunct{\mcitedefaultmidpunct}
{\mcitedefaultendpunct}{\mcitedefaultseppunct}\relax
\EndOfBibitem
\bibitem[Rosei \latin{et~al.}(2003)Rosei, Schunack, Naitoh, Jiang, Gourdon,
  Laegsgaard, Stensgaard, Joachim, and Besenbacher]{Rosei2003}
Rosei,~F.; Schunack,~M.; Naitoh,~Y.; Jiang,~P.; Gourdon,~A.; Laegsgaard,~E.;
  Stensgaard,~I.; Joachim,~C.; Besenbacher,~F. \emph{Progress in Surface
  Science}; 2003; Vol.~71; pp 95--146\relax
\mciteBstWouldAddEndPuncttrue
\mciteSetBstMidEndSepPunct{\mcitedefaultmidpunct}
{\mcitedefaultendpunct}{\mcitedefaultseppunct}\relax
\EndOfBibitem
\bibitem[Morgenstern \latin{et~al.}(2013)Morgenstern, Lorente, and
  Rieder]{Morgenstern2013}
Morgenstern,~K.; Lorente,~N.; Rieder,~K.-H. \emph{Physica Status Solidi (B)};
  2013; Vol. 250; pp 1671--1751\relax
\mciteBstWouldAddEndPuncttrue
\mciteSetBstMidEndSepPunct{\mcitedefaultmidpunct}
{\mcitedefaultendpunct}{\mcitedefaultseppunct}\relax
\EndOfBibitem
\bibitem[Eisenhut \latin{et~al.}(2018)Eisenhut, Meyer, Kr{\"{u}}ger, Ohmann,
  Cuniberti, and Moresco]{Eisenhut2018}
Eisenhut,~F.; Meyer,~J.; Kr{\"{u}}ger,~J.; Ohmann,~R.; Cuniberti,~G.;
  Moresco,~F. {Inducing the controlled rotation of single o-MeO-DMBI molecules
  anchored on Au(111)}. \emph{Surface Science} \textbf{2018}, 0--1\relax
\mciteBstWouldAddEndPuncttrue
\mciteSetBstMidEndSepPunct{\mcitedefaultmidpunct}
{\mcitedefaultendpunct}{\mcitedefaultseppunct}\relax
\EndOfBibitem
\bibitem[Nickel \latin{et~al.}(2013)Nickel, Ohmann, Meyer, Grisolia, Joachim,
  Moresco, and Cuniberti]{Nickel2013}
Nickel,~A.; Ohmann,~R.; Meyer,~J.; Grisolia,~M.; Joachim,~C.; Moresco,~F.;
  Cuniberti,~G. {Moving nanostructures: Pulse-induced positioning of
  supramolecular assemblies}. \emph{ACS Nano} \textbf{2013}, \emph{7},
  191--197\relax
\mciteBstWouldAddEndPuncttrue
\mciteSetBstMidEndSepPunct{\mcitedefaultmidpunct}
{\mcitedefaultendpunct}{\mcitedefaultseppunct}\relax
\EndOfBibitem
\bibitem[Ohmann \latin{et~al.}(2015)Ohmann, Meyer, Nickel, Echeverria,
  Grisolia, Joachim, Moresco, and Cuniberti]{Ohmann2015}
Ohmann,~R.; Meyer,~J.; Nickel,~A.; Echeverria,~J.; Grisolia,~M.; Joachim,~C.;
  Moresco,~F.; Cuniberti,~G. {Supramolecular Rotor and Translator at Work:
  On-Surface Movement of Single Atoms}. \emph{ACS Nano} \textbf{2015},
  \emph{9}, 8394--8400\relax
\mciteBstWouldAddEndPuncttrue
\mciteSetBstMidEndSepPunct{\mcitedefaultmidpunct}
{\mcitedefaultendpunct}{\mcitedefaultseppunct}\relax
\EndOfBibitem
\bibitem[Simpson \latin{et~al.}(2017)Simpson, Garc{\'{i}}a-L{\'{o}}pez,
  Petermeier, Grill, and Tour]{Simpson2017}
Simpson,~G.~J.; Garc{\'{i}}a-L{\'{o}}pez,~V.; Petermeier,~P.; Grill,~L.;
  Tour,~J.~M. {How to build and race a fast nanocar}. \emph{Nature
  Nanotechnology} \textbf{2017}, \emph{12}, 604--606\relax
\mciteBstWouldAddEndPuncttrue
\mciteSetBstMidEndSepPunct{\mcitedefaultmidpunct}
{\mcitedefaultendpunct}{\mcitedefaultseppunct}\relax
\EndOfBibitem
\bibitem[Zhang and Chi(2016)Zhang, and Chi]{Zhang2016}
Zhang,~H.; Chi,~L. {Gold–Organic Hybrids: On-Surface Synthesis and
  Perspectives}. \emph{Advanced Materials} \textbf{2016}, \emph{28},
  10492--10498\relax
\mciteBstWouldAddEndPuncttrue
\mciteSetBstMidEndSepPunct{\mcitedefaultmidpunct}
{\mcitedefaultendpunct}{\mcitedefaultseppunct}\relax
\EndOfBibitem
\bibitem[Liu \latin{et~al.}(2012)Liu, Carrasco, Santra, Michaelides, Scheffler,
  and Tkatchenko]{Liu2012}
Liu,~W.; Carrasco,~J.; Santra,~B.; Michaelides,~A.; Scheffler,~M.;
  Tkatchenko,~A. {Benzene adsorbed on metals: Concerted effect of covalency and
  van der Waals bonding}. \emph{Physical Review B - Condensed Matter and
  Materials Physics} \textbf{2012}, \emph{86}, 1--6\relax
\mciteBstWouldAddEndPuncttrue
\mciteSetBstMidEndSepPunct{\mcitedefaultmidpunct}
{\mcitedefaultendpunct}{\mcitedefaultseppunct}\relax
\EndOfBibitem
\bibitem[Pek{\"{o}}z and Donadio(2016)Pek{\"{o}}z, and Donadio]{Pekoz2016}
Pek{\"{o}}z,~R.; Donadio,~D. {Effect of van der Waals interactions on the
  chemisorption and physisorption of phenol and phenoxy on metal surfaces}.
  \emph{Journal of Chemical Physics} \textbf{2016}, \emph{145}\relax
\mciteBstWouldAddEndPuncttrue
\mciteSetBstMidEndSepPunct{\mcitedefaultmidpunct}
{\mcitedefaultendpunct}{\mcitedefaultseppunct}\relax
\EndOfBibitem
\bibitem[Mart{\'{i}}nez \latin{et~al.}(2012)Mart{\'{i}}nez, Abad,
  Gonz{\'{a}}lez, Ortega, and Flores]{Martinez2012}
Mart{\'{i}}nez,~J.~I.; Abad,~E.; Gonz{\'{a}}lez,~C.; Ortega,~J.; Flores,~F.
  {Theoretical characterization of the TTF/Au (1 1 1) interface: STM imaging,
  band alignment and charging energy}. \emph{Organic Electronics: physics,
  materials, applications} \textbf{2012}, \emph{13}, 399--408\relax
\mciteBstWouldAddEndPuncttrue
\mciteSetBstMidEndSepPunct{\mcitedefaultmidpunct}
{\mcitedefaultendpunct}{\mcitedefaultseppunct}\relax
\EndOfBibitem
\bibitem[Meyer \latin{et~al.}(2004)Meyer, Lemire, Shaikhutdinov, and
  Freund]{Meyer2004}
Meyer,~R.; Lemire,~C.; Shaikhutdinov,~S.~K.; Freund,~H.~J. {Surface chemistry
  of catalysis by gold}. \emph{Gold Bulletin} \textbf{2004}, \emph{37},
  72--124\relax
\mciteBstWouldAddEndPuncttrue
\mciteSetBstMidEndSepPunct{\mcitedefaultmidpunct}
{\mcitedefaultendpunct}{\mcitedefaultseppunct}\relax
\EndOfBibitem
\bibitem[{Eisenhut, Frank} \latin{et~al.}(2016){Eisenhut, Frank}, {Durand,
  Corentin}, {Moresco, Francesca}, {Launay, Jean-Pierre}, and {Joachim,
  Christian}]{Eisenhut2016}
{Eisenhut, Frank},; {Durand, Corentin},; {Moresco, Francesca},; {Launay,
  Jean-Pierre},; {Joachim, Christian}, Training for the 1st international
  nano-car race: the Dresden molecule-vehicle. \emph{Eur. Phys. J. Appl. Phys.}
  \textbf{2016}, \emph{76}, 10001\relax
\mciteBstWouldAddEndPuncttrue
\mciteSetBstMidEndSepPunct{\mcitedefaultmidpunct}
{\mcitedefaultendpunct}{\mcitedefaultseppunct}\relax
\EndOfBibitem
\bibitem[Kresse and Furthmüller(1996)Kresse, and Furthmüller]{KRESSE199615}
Kresse,~G.; Furthmüller,~J. Efficiency of ab-initio total energy calculations
  for metals and semiconductors using a plane-wave basis set.
  \emph{Computational Materials Science} \textbf{1996}, \emph{6}, 15 --
  50\relax
\mciteBstWouldAddEndPuncttrue
\mciteSetBstMidEndSepPunct{\mcitedefaultmidpunct}
{\mcitedefaultendpunct}{\mcitedefaultseppunct}\relax
\EndOfBibitem
\bibitem[Perdew \latin{et~al.}(1996)Perdew, Burke, and Ernzerhof]{PBE}
Perdew,~J.~P.; Burke,~K.; Ernzerhof,~M. Generalized {Gradient} {Approximation}
  {Made} {Simple}. \emph{Physical Review Letters} \textbf{1996}, \emph{77},
  3865\relax
\mciteBstWouldAddEndPuncttrue
\mciteSetBstMidEndSepPunct{\mcitedefaultmidpunct}
{\mcitedefaultendpunct}{\mcitedefaultseppunct}\relax
\EndOfBibitem
\bibitem[Kresse and Joubert(1999)Kresse, and Joubert]{Kresse1999}
Kresse,~G.; Joubert,~D. From ultrasoft pseudopotentials to the projector
  augmented-wave method. \emph{Phys. Rev. B} \textbf{1999}, \emph{59},
  1758--1775\relax
\mciteBstWouldAddEndPuncttrue
\mciteSetBstMidEndSepPunct{\mcitedefaultmidpunct}
{\mcitedefaultendpunct}{\mcitedefaultseppunct}\relax
\EndOfBibitem
\bibitem[Tkatchenko and Scheffler(2009)Tkatchenko, and
  Scheffler]{Tkatchenko2009}
Tkatchenko,~A.; Scheffler,~M. Accurate Molecular Van Der Waals Interactions
  from Ground-State Electron Density and Free-Atom Reference Data. \emph{Phys.
  Rev. Lett.} \textbf{2009}, \emph{102}, 073005\relax
\mciteBstWouldAddEndPuncttrue
\mciteSetBstMidEndSepPunct{\mcitedefaultmidpunct}
{\mcitedefaultendpunct}{\mcitedefaultseppunct}\relax
\EndOfBibitem
\bibitem[Maurer \latin{et~al.}(2016)Maurer, Ruiz, Camarillo-Cisneros, Liu,
  Ferri, Reuter, and Tkatchenko]{Tkatchenko_2016}
Maurer,~R.~J.; Ruiz,~V.~G.; Camarillo-Cisneros,~J.; Liu,~W.; Ferri,~N.;
  Reuter,~K.; Tkatchenko,~A. Adsorption structures and energetics of molecules
  on metal surfaces: Bridging experiment and theory. \emph{Progress in Surface
  Science} \textbf{2016}, \emph{91}, 72 -- 100\relax
\mciteBstWouldAddEndPuncttrue
\mciteSetBstMidEndSepPunct{\mcitedefaultmidpunct}
{\mcitedefaultendpunct}{\mcitedefaultseppunct}\relax
\EndOfBibitem
\bibitem[Scarbath-Evers \latin{et~al.}(2019)Scarbath-Evers,
  Todorovi\ifmmode~\acute{c}\else \'{c}\fi{}, Golze, Hammer, Widdra,
  Sebastiani, and Rinke]{Laura}
Scarbath-Evers,~L.~K.; Todorovi\ifmmode~\acute{c}\else \'{c}\fi{},~M.;
  Golze,~D.; Hammer,~R.; Widdra,~W.; Sebastiani,~D.; Rinke,~P. Gold diggers:
  Altered reconstruction of the gold surface by physisorbed aromatic oligomers.
  \emph{Phys. Rev. Materials} \textbf{2019}, \emph{3}, 011601\relax
\mciteBstWouldAddEndPuncttrue
\mciteSetBstMidEndSepPunct{\mcitedefaultmidpunct}
{\mcitedefaultendpunct}{\mcitedefaultseppunct}\relax
\EndOfBibitem
\bibitem[Monkhorst and Pack(1976)Monkhorst, and Pack]{Monk}
Monkhorst,~H.~J.; Pack,~J.~D. Special points for Brillouin-zone integrations.
  \emph{Phys. Rev. B} \textbf{1976}, \emph{13}, 5188--5192\relax
\mciteBstWouldAddEndPuncttrue
\mciteSetBstMidEndSepPunct{\mcitedefaultmidpunct}
{\mcitedefaultendpunct}{\mcitedefaultseppunct}\relax
\EndOfBibitem
\bibitem[J\'onsson \latin{et~al.}(1998)J\'onsson, Mills, and Jacobsen]{NEB}
J\'onsson,~H.; Mills,~G.; Jacobsen,~K. In \emph{Classical and {Quantum}
  {Dynamics} in {Condensed} {Phase} {Simulations}}; Berne,~B., Ciccotti,~G.,
  Jacobsen,~K., Eds.; World Scientific: Singapore, 1998; p 385\relax
\mciteBstWouldAddEndPuncttrue
\mciteSetBstMidEndSepPunct{\mcitedefaultmidpunct}
{\mcitedefaultendpunct}{\mcitedefaultseppunct}\relax
\EndOfBibitem
\bibitem[Johansson and Olsen(2008)Johansson, and Olsen]{biphenyl}
Johansson,~M.~P.; Olsen,~J. Torsional Barriers and Equilibrium Angle of
  Biphenyl: Reconciling Theory with Experiment. \emph{Journal of Chemical
  Theory and Computation} \textbf{2008}, \emph{4}, 1460--1471, PMID:
  26621432\relax
\mciteBstWouldAddEndPuncttrue
\mciteSetBstMidEndSepPunct{\mcitedefaultmidpunct}
{\mcitedefaultendpunct}{\mcitedefaultseppunct}\relax
\EndOfBibitem
\bibitem[Srishailam \latin{et~al.}(2019)Srishailam, Reddy, and Rao]{SRISHAILAM}
Srishailam,~K.; Reddy,~B.~V.; Rao,~G.~R. Investigation of torsional potentials,
  hindered rotation, molecular structure and vibrational properties of some
  biphenyl carboxaldehydes using spectroscopic techniques and density
  functional formalism. \emph{Journal of Molecular Structure} \textbf{2019},
  \emph{1196}, 139 -- 161\relax
\mciteBstWouldAddEndPuncttrue
\mciteSetBstMidEndSepPunct{\mcitedefaultmidpunct}
{\mcitedefaultendpunct}{\mcitedefaultseppunct}\relax
\EndOfBibitem
\bibitem[Henkelman \latin{et~al.}(2006)Henkelman, Arnaldsson, and
  J{\'{o}}nsson]{Bader}
Henkelman,~G.; Arnaldsson,~A.; J{\'{o}}nsson,~H. {A fast and robust algorithm
  for Bader decomposition of charge density}. \emph{Computational Materials
  Science} \textbf{2006}, \emph{36}, 354--360\relax
\mciteBstWouldAddEndPuncttrue
\mciteSetBstMidEndSepPunct{\mcitedefaultmidpunct}
{\mcitedefaultendpunct}{\mcitedefaultseppunct}\relax
\EndOfBibitem
\bibitem[Johnson \latin{et~al.}(2010)Johnson, Keinan, Mori-Sánchez,
  Contreras-García, Cohen, and Yang]{Johnson2010}
Johnson,~E.~R.; Keinan,~S.; Mori-Sánchez,~P.; Contreras-García,~J.;
  Cohen,~A.~J.; Yang,~W. Revealing {Noncovalent} {Interactions}. \emph{Journal
  of the American Chemical Society} \textbf{2010}, \emph{132}, 6498--6506\relax
\mciteBstWouldAddEndPuncttrue
\mciteSetBstMidEndSepPunct{\mcitedefaultmidpunct}
{\mcitedefaultendpunct}{\mcitedefaultseppunct}\relax
\EndOfBibitem
\bibitem[Otero-de-la Roza \latin{et~al.}(2012)Otero-de-la Roza, Johnson, and
  Contreras-García]{Otero2012}
Otero-de-la Roza,~A.; Johnson,~E.~R.; Contreras-García,~J. Revealing
  non-covalent interactions in solids: {NCI} plots revisited. \emph{Physical
  Chemistry Chemical Physics} \textbf{2012}, \emph{14}, 12165--12172\relax
\mciteBstWouldAddEndPuncttrue
\mciteSetBstMidEndSepPunct{\mcitedefaultmidpunct}
{\mcitedefaultendpunct}{\mcitedefaultseppunct}\relax
\EndOfBibitem
\bibitem[Boto \latin{et~al.}(2015)Boto, Contreras-García, and
  Calatayud]{Boto2015}
Boto,~R.~A.; Contreras-García,~J.; Calatayud,~M. The role of dispersion forces
  in metal-supported self-assembled monolayers. \emph{Computational and
  Theoretical Chemistry} \textbf{2015}, \emph{1053}, 322--327\relax
\mciteBstWouldAddEndPuncttrue
\mciteSetBstMidEndSepPunct{\mcitedefaultmidpunct}
{\mcitedefaultendpunct}{\mcitedefaultseppunct}\relax
\EndOfBibitem
\end{mcitethebibliography}

\end{document}